\documentclass[twocolumn,letterpaper,aps,prl,longbibliography,superscriptaddress,showpacs,floatfix]{revtex4-1}

\usepackage{graphicx}   
\usepackage{xspace} 
\usepackage{amsmath}    

\newcommand{\pt}{\mbox{$p_T$}\xspace}

\newcommand{\sqs}{\mbox{$\sqrt{s}$}\xspace}
\newcommand{\sqsn}{\mbox{$\sqrt{s_{_{NN}}}$}\xspace}
\newcommand{\heau}{\mbox{$^{3}$He$+$Au}\xspace}
\newcommand{\dau}{\mbox{$d$$+$Au}\xspace}
\newcommand{\pp}{\mbox{$p$$+$$p$}\xspace}
\newcommand{\pau}{\mbox{$p$$+$Au}\xspace}
\newcommand{\ppb}{\mbox{$p$$+$Pb}\xspace}

\begin{document}

\title{
Measurements of elliptic and triangular flow in high-multiplicity 
$^{3}$He$+$Au collisions at $\sqrt{s_{_{NN}}}=200$~GeV }

\newcommand{\abilene}{Abilene Christian University, Abilene, Texas 79699, USA}
\newcommand{\acadsin}{Institute of Physics, Academia Sinica, Taipei 11529, Taiwan}
\newcommand{\augie}{Department of Physics, Augustana College, Sioux Falls, South Dakota 57197, USA}
\newcommand{\banaras}{Department of Physics, Banaras Hindu University, Varanasi 221005, India}
\newcommand{\barc}{Bhabha Atomic Research Centre, Bombay 400 085, India}
\newcommand{\baruch}{Baruch College, City University of New York, New York, New York, 10010 USA}
\newcommand{\bnlcoll}{Collider-Accelerator Department, Brookhaven National Laboratory, Upton, New York 11973-5000, USA}
\newcommand{\bnlphys}{Physics Department, Brookhaven National Laboratory, Upton, New York 11973-5000, USA}
\newcommand{\caucr}{University of California-Riverside, Riverside, California 92521, USA}
\newcommand{\charlesczech}{Charles University, Ovocn\'{y} trh 5, Praha 1, 116 36, Prague, Czech Republic}
\newcommand{\chonbuk}{Chonbuk National University, Jeonju, 561-756, Korea}
\newcommand{\ciae}{Science and Technology on Nuclear Data Laboratory, China Institute of Atomic Energy, Beijing 102413, P.~R.~China}
\newcommand{\cns}{Center for Nuclear Study, Graduate School of Science, University of Tokyo, 7-3-1 Hongo, Bunkyo, Tokyo 113-0033, Japan}
\newcommand{\colorado}{University of Colorado, Boulder, Colorado 80309, USA}
\newcommand{\columbia}{Columbia University, New York, New York 10027 and Nevis Laboratories, Irvington, New York 10533, USA}
\newcommand{\czechtech}{Czech Technical University, Zikova 4, 166 36 Prague 6, Czech Republic}
\newcommand{\dapnia}{Dapnia, CEA Saclay, F-91191, Gif-sur-Yvette, France}
\newcommand{\debrecen}{Debrecen University, H-4010 Debrecen, Egyetem t{\'e}r 1, Hungary}
\newcommand{\elte}{ELTE, E{\"o}tv{\"o}s Lor{\'a}nd University, H-1117 Budapest, P{\'a}zm{\'a}ny P.~s.~1/A, Hungary}
\newcommand{\ewha}{Ewha Womans University, Seoul 120-750, Korea}
\newcommand{\fit}{Florida Institute of Technology, Melbourne, Florida 32901, USA}
\newcommand{\fsu}{Florida State University, Tallahassee, Florida 32306, USA}
\newcommand{\gsu}{Georgia State University, Atlanta, Georgia 30303, USA}
\newcommand{\hanyang}{Hanyang University, Seoul 133-792, Korea}
\newcommand{\hiroshima}{Hiroshima University, Kagamiyama, Higashi-Hiroshima 739-8526, Japan}
\newcommand{\howard}{Department of Physics and Astronomy, Howard University, Washington, DC 20059, USA}
\newcommand{\ihepprot}{IHEP Protvino, State Research Center of Russian Federation, Institute for High Energy Physics, Protvino, 142281, Russia}
\newcommand{\illuiuc}{University of Illinois at Urbana-Champaign, Urbana, Illinois 61801, USA}
\newcommand{\inrras}{Institute for Nuclear Research of the Russian Academy of Sciences, prospekt 60-letiya Oktyabrya 7a, Moscow 117312, Russia}
\newcommand{\instpasczech}{Institute of Physics, Academy of Sciences of the Czech Republic, Na Slovance 2, 182 21 Prague 8, Czech Republic}
\newcommand{\isu}{Iowa State University, Ames, Iowa 50011, USA}
\newcommand{\jaea}{Advanced Science Research Center, Japan Atomic Energy Agency, 2-4 Shirakata Shirane, Tokai-mura, Naka-gun, Ibaraki-ken 319-1195, Japan}
\newcommand{\jinrdubna}{Joint Institute for Nuclear Research, 141980 Dubna, Moscow Region, Russia}
\newcommand{\jyvaskyla}{Helsinki Institute of Physics and University of Jyv{\"a}skyl{\"a}, P.O.Box 35, FI-40014 Jyv{\"a}skyl{\"a}, Finland}
\newcommand{\kek}{KEK, High Energy Accelerator Research Organization, Tsukuba, Ibaraki 305-0801, Japan}
\newcommand{\korea}{Korea University, Seoul, 136-701, Korea}
\newcommand{\kurchatov}{Russian Research Center ``Kurchatov Institute", Moscow, 123098 Russia}
\newcommand{\kyoto}{Kyoto University, Kyoto 606-8502, Japan}
\newcommand{\labllr}{Laboratoire Leprince-Ringuet, Ecole Polytechnique, CNRS-IN2P3, Route de Saclay, F-91128, Palaiseau, France}
\newcommand{\lahorelums}{Physics Department, Lahore University of Management Sciences, Lahore 54792, Pakistan}
\newcommand{\lawllnl}{Lawrence Livermore National Laboratory, Livermore, California 94550, USA}
\newcommand{\losalamos}{Los Alamos National Laboratory, Los Alamos, New Mexico 87545, USA}
\newcommand{\lpc}{LPC, Universit{\'e} Blaise Pascal, CNRS-IN2P3, Clermont-Fd, 63177 Aubiere Cedex, France}
\newcommand{\lund}{Department of Physics, Lund University, Box 118, SE-221 00 Lund, Sweden}
\newcommand{\maryland}{University of Maryland, College Park, Maryland 20742, USA}
\newcommand{\mass}{Department of Physics, University of Massachusetts, Amherst, Massachusetts 01003-9337, USA}
\newcommand{\michigan}{Department of Physics, University of Michigan, Ann Arbor, Michigan 48109-1040, USA}
\newcommand{\muenster}{Institut f\"ur Kernphysik, University of Muenster, D-48149 Muenster, Germany}
\newcommand{\muhlenberg}{Muhlenberg College, Allentown, Pennsylvania 18104-5586, USA}
\newcommand{\myongji}{Myongji University, Yongin, Kyonggido 449-728, Korea}
\newcommand{\nagasaki}{Nagasaki Institute of Applied Science, Nagasaki-shi, Nagasaki 851-0193, Japan}
\newcommand{\natmephi}{National Research Nuclear University, MEPhI, Moscow Engineering Physics Institute, Moscow, 115409, Russia}
\newcommand{\newmex}{University of New Mexico, Albuquerque, New Mexico 87131, USA}
\newcommand{\nmsu}{New Mexico State University, Las Cruces, New Mexico 88003, USA}
\newcommand{\ohio}{Department of Physics and Astronomy, Ohio University, Athens, Ohio 45701, USA}
\newcommand{\ornl}{Oak Ridge National Laboratory, Oak Ridge, Tennessee 37831, USA}
\newcommand{\orsay}{IPN-Orsay, Universite Paris Sud, CNRS-IN2P3, BP1, F-91406, Orsay, France}
\newcommand{\peking}{Peking University, Beijing 100871, P.~R.~China}
\newcommand{\pnpi}{PNPI, Petersburg Nuclear Physics Institute, Gatchina, Leningrad region, 188300, Russia}
\newcommand{\riken}{RIKEN Nishina Center for Accelerator-Based Science, Wako, Saitama 351-0198, Japan}
\newcommand{\rikjrbrc}{RIKEN BNL Research Center, Brookhaven National Laboratory, Upton, New York 11973-5000, USA}
\newcommand{\rikkyo}{Physics Department, Rikkyo University, 3-34-1 Nishi-Ikebukuro, Toshima, Tokyo 171-8501, Japan}
\newcommand{\saispbstu}{Saint Petersburg State Polytechnic University, St.~Petersburg, 195251 Russia}
\newcommand{\saopaulo}{Universidade de S{\~a}o Paulo, Instituto de F\'{\i}sica, Caixa Postal 66318, S{\~a}o Paulo CEP05315-970, Brazil}
\newcommand{\seoulnat}{Department of Physics and Astronomy, Seoul National University, Seoul 151-742, Korea}
\newcommand{\stonybrkc}{Chemistry Department, Stony Brook University, SUNY, Stony Brook, New York 11794-3400, USA}
\newcommand{\stonycrkp}{Department of Physics and Astronomy, Stony Brook University, SUNY, Stony Brook, New York 11794-3800, USA}
\newcommand{\subatech}{SUBATECH (Ecole des Mines de Nantes, CNRS-IN2P3, Universit{\'e} de Nantes) BP 20722-44307, Nantes, France}
\newcommand{\tenn}{University of Tennessee, Knoxville, Tennessee 37996, USA}
\newcommand{\titech}{Department of Physics, Tokyo Institute of Technology, Oh-okayama, Meguro, Tokyo 152-8551, Japan}
\newcommand{\tsukuba}{Institute of Physics, University of Tsukuba, Tsukuba, Ibaraki 305, Japan}
\newcommand{\vandy}{Vanderbilt University, Nashville, Tennessee 37235, USA}
\newcommand{\waseda}{Waseda University, Advanced Research Institute for Science and Engineering, 17  Kikui-cho, Shinjuku-ku, Tokyo 162-0044, Japan}
\newcommand{\weizmann}{Weizmann Institute, Rehovot 76100, Israel}
\newcommand{\wigner}{Institute for Particle and Nuclear Physics, Wigner Research Centre for Physics, Hungarian Academy of Sciences (Wigner RCP, RMKI) H-1525 Budapest 114, POBox 49, Budapest, Hungary}
\newcommand{\yonsei}{Yonsei University, IPAP, Seoul 120-749, Korea}
\newcommand{\zagreb}{University of Zagreb, Faculty of Science, Department of Physics, Bijeni\v{c}ka 32, HR-10002 Zagreb, Croatia}
\affiliation{\abilene}
\affiliation{\acadsin}
\affiliation{\augie}
\affiliation{\banaras}
\affiliation{\barc}
\affiliation{\baruch}
\affiliation{\bnlcoll}
\affiliation{\bnlphys}
\affiliation{\caucr}
\affiliation{\charlesczech}
\affiliation{\chonbuk}
\affiliation{\ciae}
\affiliation{\cns}
\affiliation{\colorado}
\affiliation{\columbia}
\affiliation{\czechtech}
\affiliation{\dapnia}
\affiliation{\debrecen}
\affiliation{\elte}
\affiliation{\ewha}
\affiliation{\fit}
\affiliation{\fsu}
\affiliation{\gsu}
\affiliation{\hanyang}
\affiliation{\hiroshima}
\affiliation{\howard}
\affiliation{\ihepprot}
\affiliation{\illuiuc}
\affiliation{\inrras}
\affiliation{\instpasczech}
\affiliation{\isu}
\affiliation{\jaea}
\affiliation{\jinrdubna}
\affiliation{\jyvaskyla}
\affiliation{\kek}
\affiliation{\korea}
\affiliation{\kurchatov}
\affiliation{\kyoto}
\affiliation{\labllr}
\affiliation{\lahorelums}
\affiliation{\lawllnl}
\affiliation{\losalamos}
\affiliation{\lpc}
\affiliation{\lund}
\affiliation{\maryland}
\affiliation{\mass}
\affiliation{\michigan}
\affiliation{\muenster}
\affiliation{\muhlenberg}
\affiliation{\myongji}
\affiliation{\nagasaki}
\affiliation{\natmephi}
\affiliation{\newmex}
\affiliation{\nmsu}
\affiliation{\ohio}
\affiliation{\ornl}
\affiliation{\orsay}
\affiliation{\peking}
\affiliation{\pnpi}
\affiliation{\riken}
\affiliation{\rikjrbrc}
\affiliation{\rikkyo}
\affiliation{\saispbstu}
\affiliation{\saopaulo}
\affiliation{\seoulnat}
\affiliation{\stonybrkc}
\affiliation{\stonycrkp}
\affiliation{\subatech}
\affiliation{\tenn}
\affiliation{\titech}
\affiliation{\tsukuba}
\affiliation{\vandy}
\affiliation{\waseda}
\affiliation{\weizmann}
\affiliation{\wigner}
\affiliation{\yonsei}
\affiliation{\zagreb}
\author{A.~Adare} \affiliation{\colorado} 
\author{S.~Afanasiev} \affiliation{\jinrdubna} 
\author{C.~Aidala} \affiliation{\columbia} \affiliation{\losalamos} \affiliation{\mass} \affiliation{\michigan} 
\author{N.N.~Ajitanand} \affiliation{\stonybrkc} 
\author{Y.~Akiba} \affiliation{\riken} \affiliation{\rikjrbrc} 
\author{R.~Akimoto} \affiliation{\cns} 
\author{H.~Al-Bataineh} \affiliation{\nmsu} 
\author{J.~Alexander} \affiliation{\stonybrkc} 
\author{M.~Alfred} \affiliation{\howard} 
\author{H.~Al-Ta'ani} \affiliation{\nmsu} 
% ...
\collaboration{PHENIX Collaboration} \noaffiliation

\date{\today}



\begin{abstract}

We present the first measurement of elliptic ($v_2$) and triangular 
($v_3$) flow in high-multiplicity $^{3}$He$+$Au collisions at 
$\sqrt{s_{_{NN}}}=200$~GeV. Two-particle correlations, where the 
particles have a large separation in pseudorapidity, are compared in 
$^{3}$He$+$Au and in $p$$+$$p$ collisions and indicate that collective 
effects dominate the second and third Fourier components for the 
correlations observed in the $^{3}$He$+$Au system. The collective 
behavior is quantified in terms of elliptic $v_2$ and triangular $v_3$ 
anisotropy coefficients measured with respect to their corresponding 
event planes. The $v_2$ values are comparable to those previously 
measured in $d$$+$Au collisions at the same nucleon-nucleon 
center-of-mass energy.  Comparison with various theoretical predictions 
are made, including to models where the hot spots created by the 
impact of the three $^{3}$He nucleons on the Au nucleus expand 
hydrodynamically to generate the triangular flow.  The agreement of 
these models with data may indicate the formation of low-viscosity 
quark-gluon plasma even in these small collision systems.

\end{abstract}

\pacs{25.75.Dw}

\maketitle



The study of high-energy heavy ion collisions at the Relativistic Heavy 
Ion Collider (RHIC) and the Large Hadron Collider has produced abundant 
evidence for the formation of Quark-Gluon Plasma (QGP). Observation of 
strong elliptic and triangular flow in these $A$+$A$ collisions indicates 
that the QGP has very small viscosity and behaves like a nearly perfect 
fluid~\cite{Heinz:2013th,Adcox:2004mh,Adams:2005dq}. Recently, 
measurements at the large hadron collider in very high multiplicity events 
from collisions of \pp and \ppb have revealed similar particle emission 
patterns~\cite{Khachatryan:2010gv,Chatrchyan:2013nka,Abelev:2013wsa,Abelev:2012ola,AaD:2014lta}. 
Such features have also been detected in \dau collisions at 
RHIC~\cite{Adamczyk:2015xjc,Adare:2014keg,Adare:2013piz}. Explanations of 
the data in terms of the formation of small droplets of QGP which then 
expand hydrodynamically~\cite{Bozek:2011if} compete with alternatives 
involving novel initial-state effects (e.g. glasma 
models~\cite{Dusling:2013oia}). RHIC is uniquely suited to test these 
competing theories by its ability to engineer the size and shape of the 
initial reaction zone through collisions of \pau, \dau, and \heau, as 
proposed in Ref.~\cite{Nagle:2013lja}. In the case when small QGP droplets 
are formed, the latter two systems should have the strongest elliptic and 
triangular flow patterns, respectively.


The azimuthal anisotropy of produced particles can be quantified by the 
Fourier coefficients $v_n$ in the expansion of the particles' distribution 
as: $dN/d\phi \propto 1 + \sum_{n=1} 2 v_{n} \cos 
(n(\phi-\Psi_{n}))$~\cite{Ollitrault:1992bk}, where $n$ is the order of 
the harmonic, $\phi$ is the azimuthal angle of particles of a given type, 
and $\Psi_n$ is the azimuthal angle of the $n$th-order event plane.
In this Letter, the elliptic ($v_2$) and triangular ($v_3$) flow for 
inclusive charged hadrons produced at midrapidity $|\eta|<0.35$ in 
high-multiplicity \heau collisions at \sqsn~=~200~GeV are measured in the 
PHENIX experiment with respect to $\Psi_2$ and $\Psi_3$ event planes. The 
signatures of collective motion beyond known nonflow correlation effects 
(e.g. jets, resonances, etc.) are also examined
by comparing long-range correlations between \heau and \pp collisions.


A full description of the PHENIX experimental setup is given in 
Ref.~\cite{Adcox:2003zm}. Charged particles are reconstructed in the two 
PHENIX central-arm tracking systems comprising drift chambers and 
multi-wire proportional pad chambers (PC)~\cite{Adcox:2003zp}. Each arm 
covers ${\pi}/2$ in azimuth and $|\eta|<0.35$. The drift-chamber tracks 
are matched to hits in the outermost PC layer (PC3), reducing the 
contribution of tracks originating from decays and photon conversions.


The observed event-plane angles $\Psi_{n}^{{\rm Obs}}$ are measured in 
different pseudorapidity ranges by the beam-beam 
counters~\cite{Adare:2013nff} (BBC) and 
forward-silicon-vertex~\cite{Aidala:2013vna} (FVTX) detectors.  The PHENIX 
experiment has two BBCs, each comprising 64 quartz \v{C}erenkov radiators 
read out by photomultiplier tubes (PMTs), subtending pseudorapidity 
($3.0<|\eta|<3.9$). The FVTX~\cite{Aidala:2013vna} detector comprises two 
identical endcap assemblies, located symmetrically in the north and south 
directions. Charged particles can be detected with a high efficiency($> 
95\%$) using a cluster of mini-strip hits. The event planes are measured 
by the BBC in the Au-going (south) direction (BBC-S), which covers 
$-3.9<\eta<-3.0$, and by the reconstructed clusters in the FVTX in the 
Au-going (south) endcap (FVTX-S), which covers $-3.0<\eta<-1.0$.


The \heau data for this analysis were obtained in the 2014 run of the 
PHENIX experiment and include 1.6 billion minimum-bias (MB) triggered 
events and 480 million high-multiplicity (HM) triggered events. 
The MB trigger is defined as a 
coincidence between the north and south BBCs requiring one or more 
photomultiplier tubes firing in each, capturing 88$\pm$4\% of the total 
inelastic \heau cross section. The HM trigger is 
based on the MB trigger, but additionally requires more than 48 
photomultiplier tubes firing in the BBC-S.

\begin{figure}[tbh]
\includegraphics[width=1.08\linewidth]{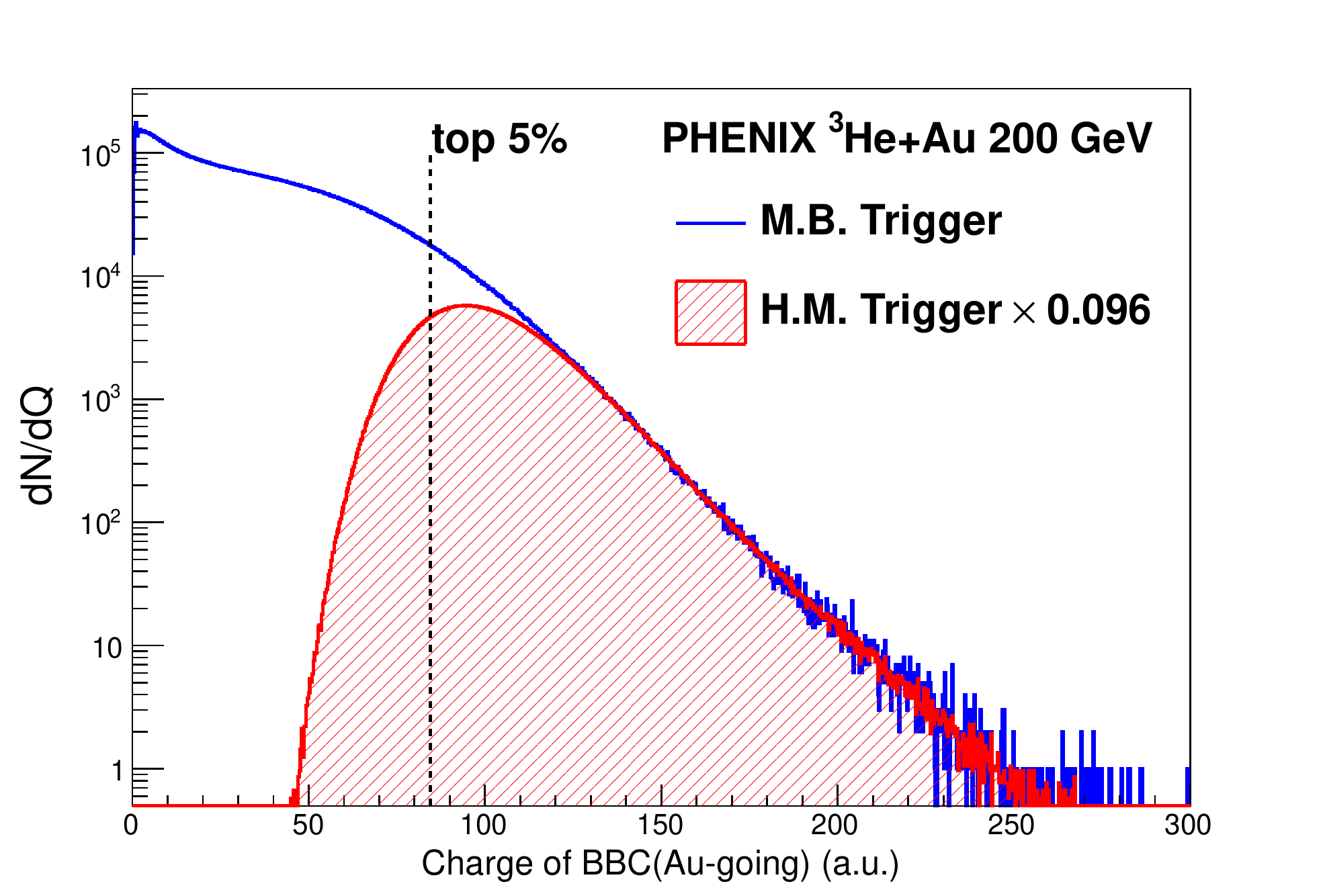}
\caption{(Color online) 
The BBC-S charge distribution in minimum bias \heau events and in 
high-multiplicity triggered events, scaled appropriately. The dashed line 
indicates the threshold selecting the 0\%--5\% most central events.
}
\label{fig:figure1}
\end{figure}

The event centrality class in \heau collisions is determined as a 
percentile of the total charge $\sum Q^{{\rm BBC-S}}$ measured in the 
BBC-S~\cite{Adare:2013nff,Adare:2013esx,Adare:2011sc,Adare:2013ezl}. The 
distribution of BBC-S PMT charge sum is shown in Fig.~\ref{fig:figure1} 
for both MB and HM triggered events.  The distribution for HM events has 
been scaled down by the relative online trigger prescale factor. For the 
MB sample, a threshold on the BBC-S charge sum shown in 
Fig.~\ref{fig:figure1} is applied to select the top 5\% central \heau 
collisions. In total, 400 million events have been used for the 
measurement of $v_2$ and $v_3$ including both HM and MB triggered events 
having BBC-S charge sum above the threshold.

Events with high BBC-S sum charge have been simulated with a Monte Carlo 
Glauber model~\cite{Loizides:2014vua,Miller:2007ri}, following the 
procedure detailed in Ref.~\cite{Adare:2013nff}. The average number of 
binary collisions ($N_{\rm coll}$), participants ($N_{\rm part}$) and 
initial-state eccentricities ($\varepsilon_{2}$, $\varepsilon_{3}$) were 
found to be consistent, within uncertainties, between the MB with the 
0\%--5\% central selection, and the HM events being used here.  
Note that in this Glauber calculation the spatial distribution from each 
participant is smeared with a two-dimensional Gaussian, 
$\sigma_{r}=0.4$~fm~\cite{Adare:2013nff}. The simulation results for 
central \heau and \dau are listed in Table~\ref{tab:table1} with the \dau 
values from Ref.~\cite{Adare:2013nff}.

\begin{table}[htb]
\caption{Monte Carlo Glauber characterization results.   
}
\begin{ruledtabular} \begin{tabular}{ccccc}
  System & $N_{\rm part}$ & $N_{\rm coll}$ & $\varepsilon_{2}$ & $\varepsilon_{3}$ \\
\hline
  0\%--5\% \heau & 25.0$\pm$1.6 & 26.1$\pm$2.0 & 0.50$\pm$0.02 & 0.28$\pm$0.02 \\
  0\%--5\% \dau & 17.3$\pm$1.2 & 18.1$\pm$1.2 & 0.54$\pm$0.04 & 0.19$\pm$0.01 \\
\end{tabular} \end{ruledtabular}
\label{tab:table1}
\end{table}


To estimate the contribution to the flow measurements from elementary 
processes, such as jets and resonance decay, we first examine azimuthal 
correlations in minimum bias \pp and central \heau events across a long 
range in pseudorapidity, with $|\Delta \eta| \sim 3.5$ between tracks in 
the PHENIX central arm at a given \pt and charge measured in the BBC PMTs.  
We use the BBCs for these correlation functions because the detector 
configuration and performance is uniform over many years, which enables us 
to combine data for \pp collisions at \sqs~=~200~GeV from running in 2005, 
2006, 2008, and 2009.  This results in 2.7 billion total minimum bias \pp 
events.

Using track-BBC pairs we construct the distribution over relative azimuth, 
and from that the normalized correlation function:
\begin{eqnarray}
  S(\Delta\phi,p_{T})=
  \frac{ d(w_{{\rm PMT}} N^{{\rm track}(p_{T}){\rm - PMT}}_{{\rm Same \; event}}) }{ d\Delta\phi} & & 
\label{eq31} \\
  C(\Delta\phi,p_{T}) =
          \frac{S(\Delta\phi,p_{T})}{M(\Delta\phi,p_{T})} \:
          \frac{\int_{0}^{2\pi} M(\Delta\phi,p_{T}) \, d\Delta\phi}{\int_{0}^{2\pi} S(\Delta\phi,p_{T}) \, d\Delta\phi} & & 
  \label{eq:def_corr_function} 
\end{eqnarray}
The weighting $w_{{\rm PMT}}$ for each pair is taken as the PMT charge.  
The signal distribution $S$ is over pairs in the same event;  the mixed 
distribution $M$ is over pairs from different events in the same event 
centrality and $z$ vertex bin, and serves to correct for any nonuniformity 
in acceptance over $\Delta \phi$.

\begin{figure}[tbh]
  \includegraphics[width=1.02\linewidth]{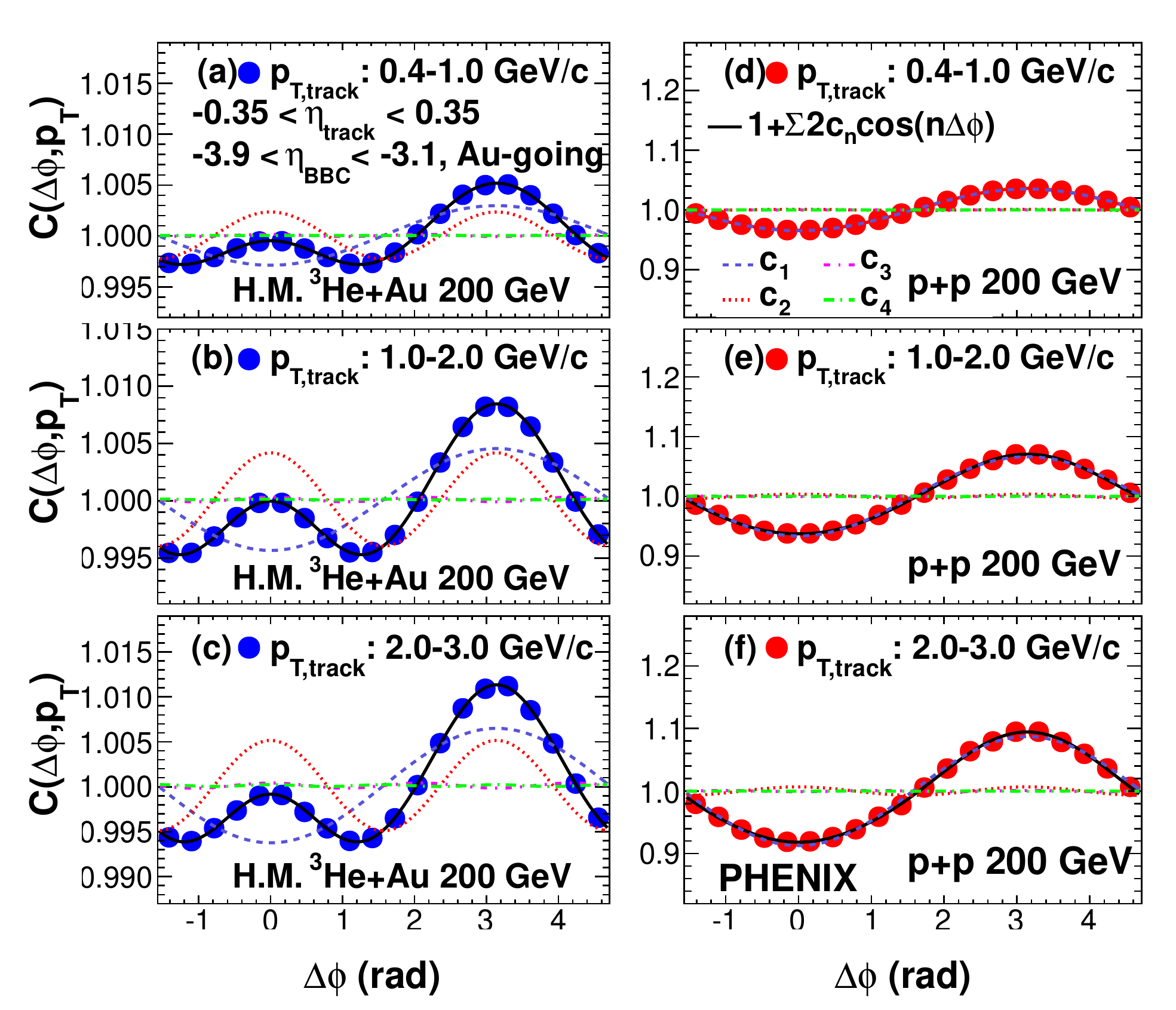}
  \caption{(Color online) 
The azimuthal correlation functions $C(\Delta\phi,p_{T})$, as defined in 
Eq.~\ref{eq:def_corr_function}, for track-BBC pairs with different 
track \pt selections in (a)--(c) HM \heau collisions and (d)--(f) minimum 
bias $p$$+$$p$ collisions both at \sqsn~=~200~GeV.  The track \pt bins are 
(a),(d) 0.4--1.0, (b),(e) 1.0--2.0, and (c),(f) 2.0--3.0~GeV/$c$. Each 
correlation function is fit with a four-term Fourier cosine expansion; the 
individual components $n=1$ to $n=4$ are drawn on each panel, together 
with the fit function sum. } 
\label{fig:corr_functions} 
\end{figure}

Figure~\ref{fig:corr_functions} shows the correlation functions 
$C(\Delta\phi,p_{T})$ for different \pt bins, for HM \heau collisions 
using (a)--(c) BBC-S and (d)--(f) minimum bias \pp collisions using both 
BBCs. We analyze these shapes by fitting each $C(\Delta\phi,p_{T})$ to 
a four-term Fourier cosine expansion, $f(\Delta\phi) = 1 + \sum_{n=1}^{4} 
\, 2c_{n}(p_{T}) \, \cos ( n \, \Delta\phi )$.  The sum function and each 
individual cosine component are plotted in Fig.~\ref{fig:corr_functions} 
for each distribution. Central \heau collisions show a clearly visible 
enhancement of near-side pairs, producing a local maximum in the 
distribution at $\Delta\phi \sim 0$. In contrast, \pp collisions can be 
described almost completely by the dipole term cos($\Delta\phi$), as 
expected generically from back-to-back jets and transverse momentum 
conservation.

\begin{figure}[htb]
  \includegraphics[width=1.13\linewidth]{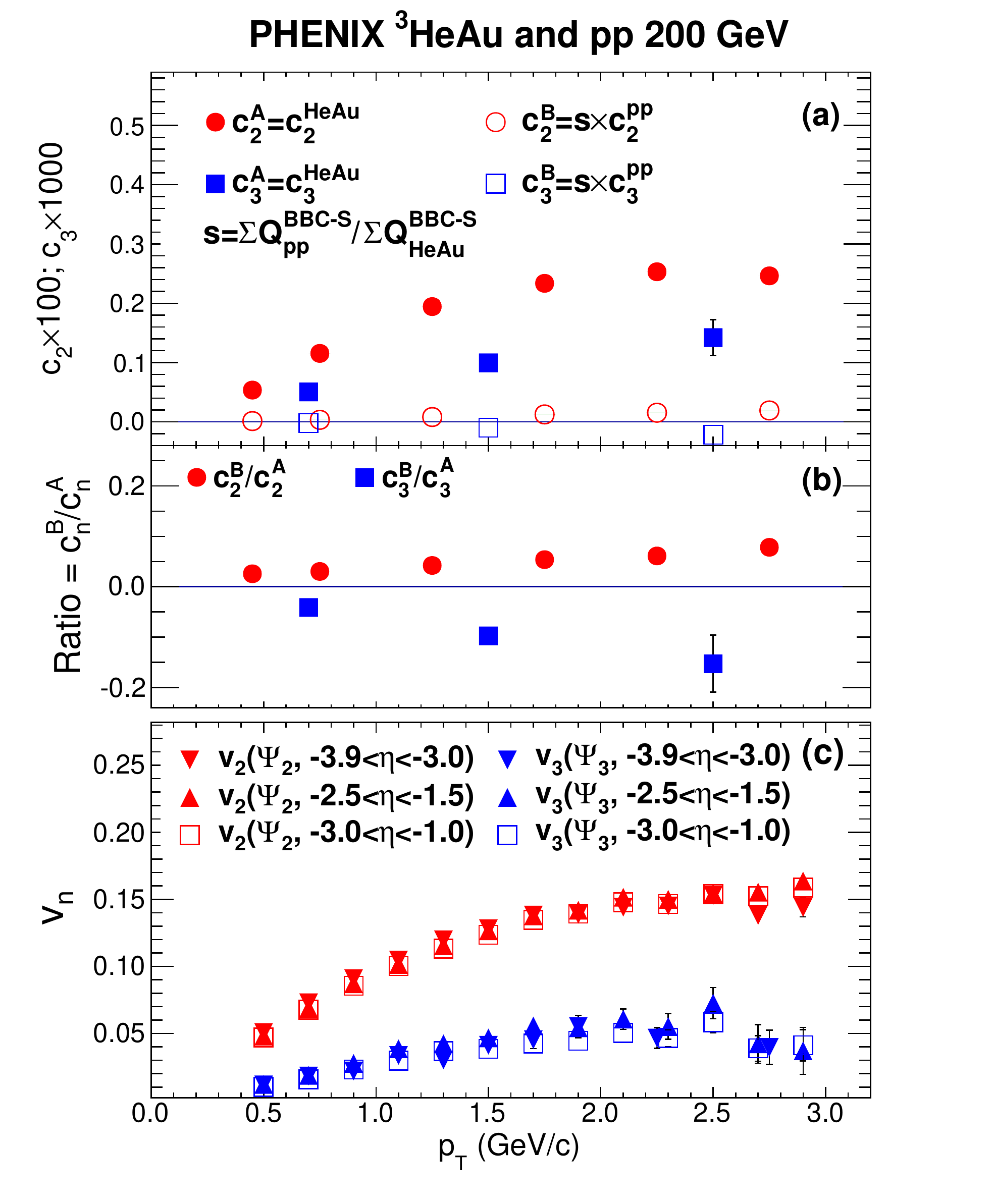}
  \caption{(Color online)
(a)~The $c_n(p_T)$ coefficients for track-BBC pairs from high multiplicity 
$^{3}$He+Au collisions (filled, denoted ``$c^{{\rm A}}$'') and 
$c_{n}(p_{T})$ for pairs in minimum bias \pp collisions times the dilution 
factor $\left( \sum Q^{{\rm BBC-S}} \right)_{p+p} / \left( \sum Q^{{\rm 
BBC-S}} \right)_{{\rm HeAu}}$ (open, denoted ``$c^{{\rm B}}$'' ). (b)~The 
ratio $c^{{\rm B}}/c^{{\rm A}}$ is shown with statistical error bars. 
(c)~Comparison of extracted values of $v_{2}$ and $v_{3}$ for midrapidity 
tracks in central \heau using event plane measurements with detectors in 
different pseudorapidity intervals (see text).
}
\label{fig:cn_vn_comparisons}
\end{figure}

As in our \dau analysis~\cite{Adare:2014keg}, we estimate
quantitatively the correlation strength that would be
observed in a class of \heau collisions purely from
elementary processes, such as in \pp collisions.
Approximating the \heau collisions as superpositions of some
number $N$ of \pp collisions, we would then expect the
correlation strengths $c_n$ from the superposition to be the
same as in \pp, but diluted by a factor of $1/N$.  We then
approximate this dilution factor as simply the ratio of
total charge observed in the BBC-S detector in \pp versus \heau:
\begin{equation}
c_{n}^{{\rm HeAu \; elementary}}(p_{T}) \simeq c_{n}^{p+p}(p_{T}) 
\frac{\left( \sum Q^{{\rm BBC-S}} \right)_{p+p}}
{\left( \sum Q^{{\rm BBC-S}} \right)_{{\rm HeAu}}
}
\end{equation}
with a numerical value of ${1/(20.6 \pm 0.4)}$. 
Figure~\ref{fig:cn_vn_comparisons}(a) shows the $c_{n}$'s from the \heau 
correlation functions, and from the \pp with the dilution factor applied.  
The $c_{2}^{pp}$ is found to be positive and the $c_{3}^{pp}$ to be 
negative. This indicates that the correlation of elementary processes 
contributes positively to the $v_2$ but negatively to the $v_3$. The 
ratios in Fig.~\ref{fig:cn_vn_comparisons}(b) show that the relative 
correlation strength in \heau from elementary processes grows with \pt, as 
might be expected from jet processes for example, but does not exceed 
7\% (15\%) for $c_{2}$ ($c_{3}$).

We now quantify the strength of collective behavior through the $v_n$ 
coefficients. The $v_n$ coefficients for charged hadrons at midrapidity 
are measured in central \heau events via the event-plane 
method~\cite{Poskanzer:1998yz} as $v_{n}(p_{T})=\langle \cos n (\phi^{{\rm 
Particle}}(p_{T}) - \Psi_{n}^{{\rm Obs}} ) \rangle /{\rm 
Res}(\Psi_{n}^{{\rm Obs}})$, where the average is over particles in the 
same \pt bin and events of the same centrality. The $n$th-order 
event-plane direction $\Psi_{n}^{{\rm Obs}}$ is determined in each event with 
the BBC-S or FVTX-S detectors.  
The $\Psi_{n}^{{\rm Obs}}$ are corrected for each detector with a standard 
event-plane flattening 
technique~\cite{Poskanzer:1998yz,Adler:2003kt,Adler:2005rg,Afanasiev:2009wq} 
to remove the effect of any small, residual nonuniformities in the 
detector response.  As a cross-check we use event planes from both the 
full FVTX-S covering $-3.0<\eta<-1.0$ and a subsection covering 
$-2.5<\eta<-1.5$. The choice of the latter is to avoid edge effects and 
still retain good FVTX-S acceptance.

We calculate the resolutions ${\rm Res}(\Psi_{n}^{{\rm Obs}})$ for each 
detector at each $n$ using the standard three-event-plane 
method~\cite{Poskanzer:1998yz,Adare:2014keg}, combining two event planes 
with the $n$th order event plane determined from central-arm tracks, 
restricted to low \pt (0.2 $<p_T<$ 2.0~GeV/$c$) to minimize contribution 
from jet fragments. In the case of the $n=2$ event plane, the resolution 
is also estimated using the first-order event plane measured with 
spectator neutrons in the shower-maximum detector of the zero-degree 
calorimeter ~\cite{Baltz:1998ex,Afanasiev:2009wq} on the Au-going side 
($\eta<-6.5$). The values for the resolution obtained in both methods are 
found to be consistent within uncertainties.

The event plane resolutions for each detector and order are shown in 
Table~\ref{tab:reso-bbc-fvtx}. The $v_2$ and $v_3$ measured using the 
three event planes described above are shown in 
Fig.~\ref{fig:cn_vn_comparisons}(c), and are consistent between detectors 
to within 5\%(15\%) for $v_2$($v_3$) over the whole \pt range.

\begin{table}[tbh]
\caption{
The resolution of $n$th-order event-plane angles measured by the BBC-S and 
FVTX-S detectors. 
}
\begin{ruledtabular} \begin{tabular}{lcc}
  Subsystem & ${\rm Res}(\Psi_{2}^{{\rm Obs}})$ & ${\rm Res}(\Psi_{3}^{{\rm Obs}})$ \\
\hline
   BBC-S (-3.9 $<\eta<$ -3.0) & 0.110 & 0.034 \\
  FVTX-S (-2.5 $<\eta<$ -1.5) & 0.232 & 0.052 \\
  FVTX-S (-3.0 $<\eta<$ -1.0) & 0.274 & 0.070 \\
\end{tabular} \end{ruledtabular}
\label{tab:reso-bbc-fvtx}
\end{table}


The main sources of systematic uncertainty for the $v_n$ measurements are: 
(1)~track backgrounds from weak decays and photon conversions; 
(2)~multiple \heau collisions in a bunch crossing (pile-up); 
(3)~biases in event plane determination; 
(4)~the effect of detector alignment and performance on the $v_n$
measurement; and 
(5)~elementary process/nonflow correlations. 
We assign the following values to account for these systematic uncertainties:
(1)~We estimate the track background contribution by reducing the spatial 
matching windows in PC3 from 3$\sigma$ to 2$\sigma$, and find a change of 
less than 2\%(5\%) fractionally in $v_{2}$($v_{3}$).
(2)~We expect the $v_n$ from pile-up events to be modestly reduced.  
Conservatively assuming that pile-up events that contaminate the sample at 
the level of 4\%--5\% have a negligible $v_n$, this results in a 
$^{+0}_{-5}$\% systematic uncertainty.
(3)~Event plane effects are estimated from $v_n$ measurements using 
different event plane detectors as shown in  
Fig.~\ref{fig:cn_vn_comparisons}(c). They are no more than 5\%(15\%) for 
$v_{2}$($v_{3}$).
(4)~The difference of $v_n$ for charged hadrons measured by the east and 
west DC arms are found to be
less than 2\%(15\%) for $v_{2}$($v_{3}$). 
(5)~The contribution from nonflow correlations at each \pt is estimated 
from Fig.~\ref{fig:cn_vn_comparisons}(b), reaching a maximum of 7\%(15\%) 
for $v_2$($v_3$).
We do not attempt to correct for this contribution and instead treat it as 
a systematic uncertainty. All of these contributions are summed in 
quadrature.
 
\begin{figure}[htbp]
  \includegraphics[width=1.08\linewidth]{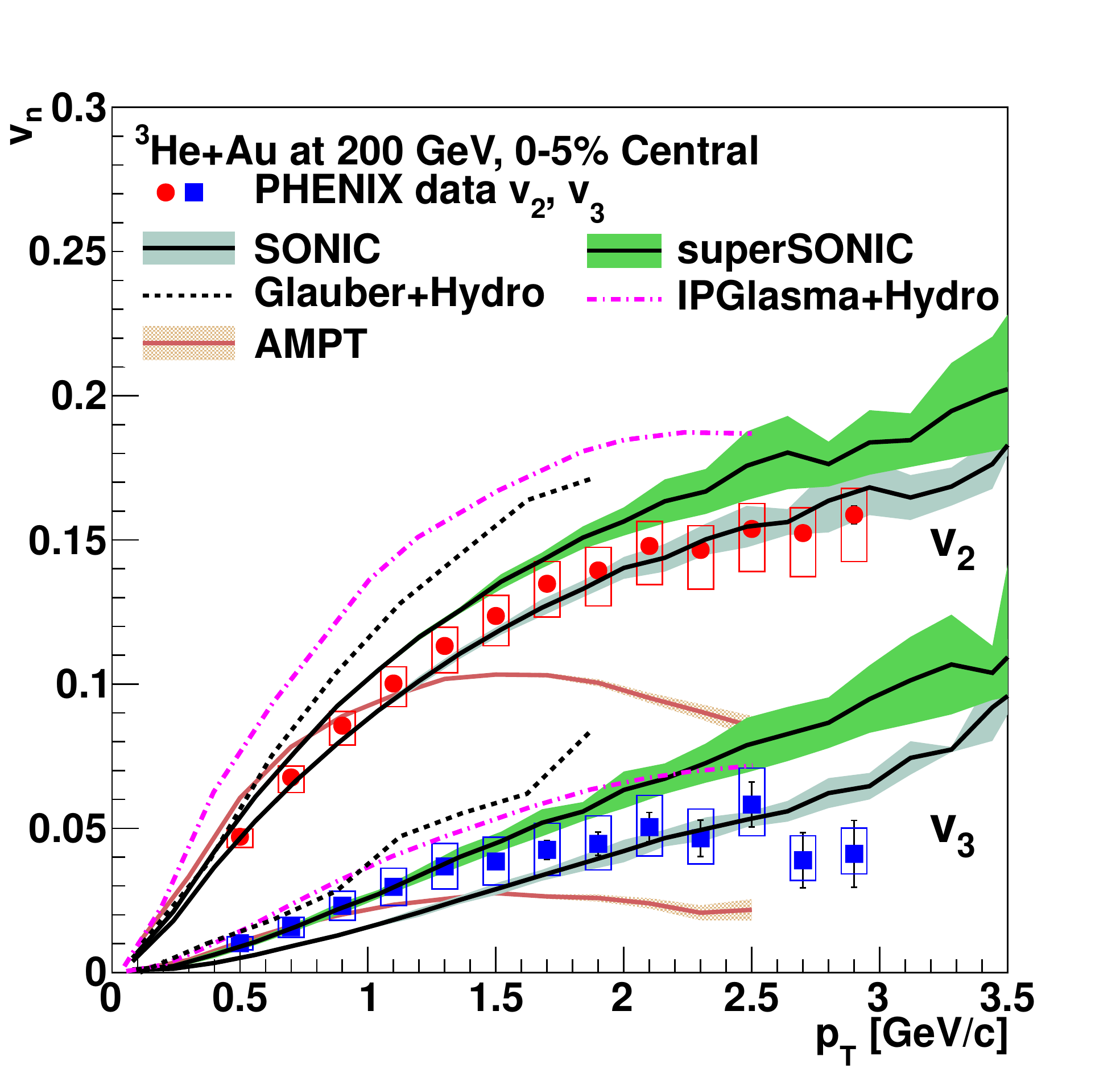}
  \caption{(Color online) 
Results for $v_{2}$ (circles) and $v_{3}$ (squares) as a function of \pt 
for inclusive charged hadrons at midrapidity in 0\%--5\% central \heau 
collisions at \sqsn~=~200~GeV; error bars are statistical and shaded bars 
are systematic uncertainties as described in the text. Also shown are 
various theoretical calculations, see text for details and references.
}
\label{fig:figure4}
\end{figure}

The final $v_n$ results are determined using the event plane measured in 
the FVTX-S covering -3.0$<\eta<$-1.0, and these are shown in 
Fig.~\ref{fig:figure4}, with the systematic uncertainties as described 
above. We observe sizable $v_{2}$ and $v_{3}$ anisotropies that both rise 
as a function of \pt. It is notable that the $v_{2}(p_{T})$ values for 
central \heau collisions are very similar within uncertainties with those 
reported earlier in central \dau collisions~\cite{Adare:2014keg}. In 
scenarios where these anisotropies reflect the initial geometry, this 
similarity would be expected as the initial eccentricities 
$\varepsilon_{2}$ for central \dau and \heau are essentially identical, as 
shown in Table~\ref{tab:table1}. The same calculations indicate a much 
larger $\varepsilon_{3}$ in \heau compared with \dau collisions.  
However, the \dau data used in \cite{Adare:2014keg} were taken in 2008, 
without a central trigger and before the FVTX was installed, and did not 
allow extraction of a statistically significant $v_{3}$ in \dau.

We now compare the experimental data with theory predictions in the 
literature. Four such predictions shown in Fig.~\ref{fig:figure4} employ 
viscous hydrodynamics with $\eta/s$ at or near the conjectured lower bound 
$1/4\pi$~\cite{Kovtun:2004de}. The Glauber+Hydro~\cite{Bozek:2015qpa} 
(IP-Glasma+Hydro~\cite{Schenke:2014gaa}) utilize Glauber (IP-Glasma) 
initial conditions, and both over predict the magnitude of the 
$v_{2}$ and $v_{3}$ data. 
Improved agreement may be achieved by utilizing a larger value of $\eta/s$ 
or by the inclusion of a transition from QGP to a hadronic cascade, which 
has much larger viscous effects and thus decreases the overall flow. The 
{\sc sonic} calculation~\cite{Nagle:2013lja} employs Glauber initial 
conditions, viscous hydrodynamics, and then at $T=170$~MeV a transition to 
a hadronic cascade.  The {\sc supersonic} 
calculation~\cite{Romatschke:2015gxa} additionally includes 
pre-equilibrium dynamics that boosts the initial velocity fields at the 
earliest times. The impact of pre-equilibrium is modest on the $v_{2}$ 
values and both calculations agree with the data within uncertainties.
The effect of pre-equilibrium on $v_{3}$ is significantly larger as the 
triangular flow takes longer to develop~\cite{Nagle:2013lja}. 
The {\sc supersonic} prediction agrees well with the experimental data for 
$p_{T} < 1.5$ GeV/$c$, and then the data trends towards the {\sc sonic} 
prediction at higher \pt. 

Lastly, we compare to calculations utilizing the A-Multi-Phase-Transport 
({\sc ampt}) model~\cite{Lin:2004en}, which incorporates both partonic and 
hadronic scattering, and has recently been compared with anisotropies in 
central \ppb and \dau collisions~\cite{Bzdak:2014dia,Koop:2015wea}. 
{\sc ampt} results for \heau agree reasonably with the 
experimental $v_2$ and $v_3$ data for $p_T<1$~GeV and then significantly 
underpredict the data.  Possible underlying causes of the anisotropies 
within the {\sc ampt} model are discussed in Ref.~\cite{He:2015hfa}.


We have presented first results on azimuthal anisotropies $v_{2}$ and 
$v_{3}$ in central \heau collisions at \sqsn~=~200~GeV.  Calculations 
including hydrodynamic expansion of the initial hot spots created in 
\heau collisions qualitatively describe the data.  Further comparison with 
different theoretical models should be informative in terms of the 
contributions from the initial geometry and each time stage in the medium 
evolution including pre-equilibrium.  Forthcoming results from \pau 
collisions at RHIC will provide a full suite of geometries with highly 
asymmetric collisions to constrain the origin of the observed anisotropies.




We thank the staff of the Collider-Accelerator and Physics
Departments at Brookhaven National Laboratory and the staff of
the other PHENIX participating institutions for their vital
contributions.  We acknowledge support from the 
Office of Nuclear Physics in the
Office of Science of the Department of Energy,
the National Science Foundation, 
a sponsored research grant from Renaissance Technologies LLC,
Abilene Christian University Research Council, 
Research Foundation of SUNY, and
Dean of the College of Arts and Sciences, Vanderbilt University 
(U.S.A),
Ministry of Education, Culture, Sports, Science, and Technology
and the Japan Society for the Promotion of Science (Japan),
Conselho Nacional de Desenvolvimento Cient\'{\i}fico e
Tecnol{\'o}gico and Funda\c c{\~a}o de Amparo {\`a} Pesquisa do
Estado de S{\~a}o Paulo (Brazil),
Natural Science Foundation of China (P.~R.~China),
Ministry of Science, Education, and Sports (Croatia),
Ministry of Education, Youth and Sports (Czech Republic),
Centre National de la Recherche Scientifique, Commissariat
{\`a} l'{\'E}nergie Atomique, and Institut National de Physique
Nucl{\'e}aire et de Physique des Particules (France),
Bundesministerium f\"ur Bildung und Forschung, Deutscher
Akademischer Austausch Dienst, and Alexander von Humboldt Stiftung (Germany),
National Science Fund, OTKA, K\'aroly R\'obert University College, 
and the Ch. Simonyi Fund (Hungary),
Department of Atomic Energy and Department of Science and Technology (India), 
Israel Science Foundation (Israel), 
Basic Science Research Program through NRF of the Ministry of Education (Korea),
Physics Department, Lahore University of Management Sciences (Pakistan),
Ministry of Education and Science, Russian Academy of Sciences,
Federal Agency of Atomic Energy (Russia),
VR and Wallenberg Foundation (Sweden), 
the U.S. Civilian Research and Development Foundation for the
Independent States of the Former Soviet Union, 
the Hungarian American Enterprise Scholarship Fund,
and the US-Israel Binational Science Foundation.


%
 
\end{document}